\documentclass[conference]{IEEEtran}
\IEEEoverridecommandlockouts
\usepackage{cite}
\usepackage{amsmath,amssymb,amsfonts}
\usepackage{algorithmic}
\usepackage{graphicx}
\usepackage{textcomp}
\usepackage{xcolor}
\usepackage{tabularx}
\usepackage{array}
\usepackage{tikz}
\usepackage[margin=0.7in]{geometry}
\usepackage{adjustbox}
\usepackage{enumitem} 
\usepackage{ragged2e}
\usepackage{balance}
\usepackage{multicol}
\usepackage{url}
\usepackage{booktabs}
\usepackage{needspace} 

\newcolumntype{P}[1]{>{\centering\arraybackslash}m{#1}}

\def\BibTeX{{\rm B\kern-.05em{\sc i\kern-.025em b}\kern-.08em
T\kern-.1667em\lower.7ex\hbox{E}\kern-.125emX}}

\makeatletter
\newcommand{\newlineauthors}{%
  \end{@IEEEauthorhalign}\hfill\mbox{}\hfill\vspace*{4mm}\\%
  \mbox{}\hfill\begin{@IEEEauthorhalign}
}
\makeatother

\begin{document}

\title{Characterizing and Codifying Malware Sophistication\\}

\author{\IEEEauthorblockN{ Angelo Porcella}
\IEEEauthorblockA{\textit{Montana State University} \\
Bozeman, USA \\
angeloporcella@montana.edu}
\and
\IEEEauthorblockN{Zachary Wadhams}
\IEEEauthorblockA{\textit{Montana State University} \\
Bozeman, USA \\
zachary.wadhams@montana.edu}
\and
\IEEEauthorblockN{Clemente Izurieta}
\IEEEauthorblockA{\textit{Montana State University} \\
Bozeman, USA \\
clemente.izurieta@montana.edu}
\newlineauthors
\IEEEauthorblockN{Jonathan Crussell}
\IEEEauthorblockA{\textit{Sandia National Laboratories} \\
Livermore, CA \\
jcrusse@sandia.gov}
\and
\IEEEauthorblockN{Ann Marie Reinhold}
\IEEEauthorblockA{\textit{Montana State University} \\
Bozeman, USA \\
reinhold@montana.edu}
}

\maketitle

\begingroup
\renewcommand\thefootnote{}
\footnotetext{%
\footnotesize
© 2026 IEEE. Personal use of this material is permitted.
Permission from IEEE must be obtained for all other uses, in any current or future media,
including reprinting/republishing this material for advertising or promotional purposes,
creating new collective works, for resale or redistribution to servers or lists,
or reuse of any copyrighted component of this work in other works.

\medskip
This is the author’s accepted version of: A. Porcella, Z. Wadhams, C. Izurieta, J. Crussell, and A. M. Reinhold, “Characterizing and Codifying Malware Sophistication,” 2026 Intermountain Engineering, Technology and Computing (IETC), Provo, UT, USA, 2026, pp. 1–6. doi: \url{https://doi.org/10.1109/IETC69527.2026.11568673}.
}
\addtocounter{footnote}{-1}
\endgroup

\begin{abstract}
``Sophisticated" is widely used to describe malware, yet it lacks a consistent definition within academic literature. While existing software quality and complexity metrics offer some insight into malware structure, they do not capture the broader adversarial and operational traits that contribute to real-world threat potential. This paper presents a systematization of existing approaches for assessing malware quality using static binary analysis. We define malware sophistication through a quality-focused lens by reinterpreting select characteristics from the ISO/IEC 25010 software quality standard, including reliability, maintainability, flexibility, and security, and evaluating their applicability to malware binaries. We identify which characteristics are both relevant to malware and measurable through static analysis, forming the basis for future frameworks that consistently assess malware sophistication when source code is unavailable or dynamic execution is infeasible
\end{abstract}


\begin{table*}[ht!]
\centering
\caption{Definitions Related to Software and Malware Sophistication}
\begin{tabularx}{\textwidth}{@{}P{3cm}X@{}}
\toprule
\textbf{Term} & \textbf{Definition} \\
\midrule
\textbf{Malware} & Software or firmware intended to perform an unauthorized process with an adverse impact on the confidentiality, integrity, or availability of a system~\cite{b3}. \\
\midrule
\textbf{Sophistication} & The amount of specialized knowledge and effort expended in the development of a piece of software to satisfy the goals and objectives of the development team. \\
\midrule
\textbf{Quality} & How well a software product conforms to its requirements and meets the needs of its users. From a software product perspective, quality refers to characteristics such as reliability, usability, performance, and security~\cite{b48}. \\
\midrule
\textbf{Novelty} & Degree in which a malware sample is unique, purpose built, or unrecognizable from pattern matching antivirus software. \\
\midrule
\textbf{Obfuscation} & The act of making software harder to discover or understand, either through manual or automated analysis, without altering the required functionality of the software~\cite{b46}. \\
\midrule
\textbf{Complexity} & The amount and predictability of emergent behaviors arising from interactions both internally (i.e., components) and externally (i.e., the larger system). Complex software has components whose interactions evolve, with potentially unbounded states.~\cite{b47}.  \\
\bottomrule
\end{tabularx}
\label{tab:malware_definitions}
\end{table*}

\section{Introduction}
The number of unique malware samples each year has been increasing at an extraordinary rate. In 2024, AV-Test.org reported an average of more than 450,000 new malicious programs registered daily~\cite{b1}. As malware becomes easier to develop and distribute, it is increasingly important to establish a systematic method for assessing the sophistication of a sample using only its binary. Such a method would help analysts better understand the capabilities and intent of adversarial groups, enabling more effective threat prioritization and resource allocation. However, this is a formidable task. Sophistication reflects a combination of characteristics, such as code complexity, obfuscation, novelty, and functionality, many of which are difficult to measure without access to source code, particularly because malware is deliberately engineered to resist analysis.

This systematization of knowledge (SoK) investigates how existing software quality metrics and other statically measurable characteristics can be applied to assess malware sophistication. Despite the growing importance of this task, no standardized framework exists for evaluating sophistication based solely on binary analysis. Most current approaches to malware analysis focus on classification or behavior-based labeling rather than quantification~\cite{b73,b74}. Notably, in 2001, the Malware Rating System proposed assessing threat levels based on behavioral criteria such as payload potential, proliferation potential, and hostility level~\cite{b7}. However, more than two decades later, this model has seen little adoption. In response, we propose a broader set of criteria that are measurable from malware binaries, supported by examples of tools that assist in quantifying sophistication and, by extension, threat level.

\section{Background}
Efforts to evaluate malware sophistication remain fragmented, in part because no standardized model currently defines how to quantify the complexity, quality, or functionality of malicious software~\cite{b75}. A consistent framework for evaluating sophistication would enable analysts to better interpret threat-actor intent and assess the potential impact of newly discovered samples. While traditional software quality metrics offer mature methodologies for assessing engineering quality, they do not translate directly to malware, particularly when analysis is limited to binaries alone. This section describes what constitutes malware sophistication, reviews how malware analysis is currently performed, examines how those methods inform notions of sophistication, and explains how software quality is measured using established standards.

\subsection{Malware Sophistication}
\looseness=-1
``Sophistication" is frequently used in threat reports and academic papers to describe advanced malware, yet the term lacks a consistent definition. Some authors use it to refer to technical complexity, while others emphasize evasion capabilities, operational stealth, or novelty. This ambiguity complicates comparative analysis and hinders the development of standardized assessment methods.

\looseness=-1
We propose that sophistication reflects the amount of specialized knowledge and effort expended in the development of malware. This definition distinguishes sophisticated samples from commodity malware, which may be functionally effective but requires little expertise to produce or deploy. Under this framing, sophistication is not simply a function of what malware does, but how well it is engineered to achieve its objectives while resisting detection and analysis.

\looseness=-1
Three major barriers complicate the measurement of malware sophistication. First, source code is rarely available. In nearly all real-world cases, analysts must examine compiled binaries without access to original development artifacts~\cite{b52}. This limitation restricts applicability of many established software quality metrics, such as maintainability indices or Halstead complexity, which rely on source-level constructs~\cite{b49}. Compounding this, malware authors actively deploy obfuscation techniques ranging from simple packers and LLVM obfuscation passes to polymorphic engines that can prevent meaningful disassembly altogether~\cite{b33}. These practices introduce the risk of a \textit{deceptive simplicity paradox}: a technically advanced sample may appear unsophisticated to current analysis methods due to deliberate evasion of widely-used tools such as Ghidra, CAPA, or YARA.

\looseness=-1
Second, malware spans a wide spectrum of goals, capabilities, and operational environments. A trojan~\cite{b71} may depend heavily on evasion while ransomware~\cite{b72} emphasizes capability and impact over stealth. Metrics meaningful for one class may not apply others, requiring sophistication to be evaluated in context. Third, no unified framework exists for evaluating malware quality. While prior work has explored quality subcharacteristics using software engineering metrics~\cite{b5}, there is no established standard tailored to malicious software.

\subsection{Current Methods of Malware Analysis}

\looseness=-1
Researchers have explored various methods for assessing malware quality and complexity, often adapting traditional software engineering metrics to the domain of malicious code. These studies assume access to malware source code or high-fidelity decompilation and apply metrics such as LOC, cyclomatic complexity, and object-oriented design indicators to evaluate maintainability, modularity, or development effort~\cite{b5,b10}. While these approaches can reveal trends in malware evolution, their reliance on source-level artifacts limits their applicability in operational contexts.

\looseness=-1
The MalSource study by Calleja et al. analyzed 456 malware samples with available source code spanning four decades, using metrics such as source lines of code, function points, and maintainability indices~\cite{b5}. Their findings show a progression from small, individually developed malware toward larger, modular systems. Similarly, Mercaldo et al. applied software metrics to decompiled Android APKs to infer increasing sophistication over time~\cite{b10}. Both studies demonstrate that traditional metrics can capture meaningful properties of malware development, but only when high-level representations are available. In practice, analysts almost always operate on compiled binaries, often in the presence of obfuscation or packing~\cite{b52}.

\looseness=-1
In the absence of source code, analysts rely primarily on static analysis—inspecting binaries without execution using techniques such as disassembly, control-flow graph extraction, and string inspection. While dynamic analysis in sandboxed environments can reveal runtime behaviors, it is resource-intensive, susceptible to evasion, and beyond the scope of this work. While some tools exist for binary-level analysis, there are far fewer in number and less mature than those designed for source code~\cite{b70,b69}.

\subsection{Sophistication Proxies}

\looseness=-1
In traditional software engineering, effort is often estimated using metrics like lines of code, function points, or development time. However, malware authors may intentionally manipulate these surface-level metrics. Dead code insertion adds non-functional instructions solely to bloat control-flow graphs, while binary padding appends junk data to exceed scanning limits or evade hash-based defenses. Both strategies are recognized in MITRE ATT\verb|&|CK{\textsuperscript{\tiny\textregistered}} as forms of defense evasion~\cite{b60}. These challenges suggest that metrics of effort derived from binaries must be interpreted cautiously—apparent complexity may reflect defensive measures rather than underlying design.

\looseness=-1
Malware threat analysis aims to assess the risk, intent, and technical characteristics of malicious software. Although its primary goal is not to measure sophistication, many of the same characteristics, such as capability, platform compatibility, and structural complexity, are relevant to understanding how advanced a sample may be. Existing threat assessment approaches provide useful insight into how observable features from binary analysis can inform evaluations of malware. 

\looseness=-1
Maasberg, Ko, and Beebe propose a structured approach to malware threat assessment that emphasizes comparability across samples~\cite{b9}. Their model draws on MITRE's Malware Attribute Enumeration and Characterization (MAEC) framework, which organizes threat information into standardized attributes. These include the objectives of the malware author, the capabilities of the malware, and its ability to operate on multiple operating systems. By framing these elements as components of threat level, the authors indirectly capture several traits that also signal sophistication, such as technical breadth and operational versatility. While their work is not focused on quality measurement directly, it demonstrates how structured attribute models can measure meaningful differences between malware samples using observable static features.

\looseness=-1
Dai et al. present a multi-stage model for assessing malware threats based on the fusion of static and dynamic analysis data~\cite{b8}. Their static analysis component focuses on extracting structural features from the binary, including disassembled code, control flow graphs, and file metadata. These features are then combined with behavioral observations to produce a composite threat score. While the primary aim of this system is to classify malware based on risk, the inclusion of control-flow complexity and structural analysis highlights how low-level features can serve as indirect indicators of sophistication. The model underscores the potential of binary-derived metrics to inform broader evaluations, even in the absence of source code or execution traces.

Although existing threat assessment models focus primarily on risk, they demonstrate that static features such as control flow, compatibility, and functional scope can infer meaningful differences between samples, providing a foundation for more targeted models of malware quality.

\subsection{Software Quality Measurement}
\looseness=-1
The ISO/IEC 25010 standard defines a set of characteristics used to evaluate both software product quality and quality in use~\cite{b6}. It is widely adopted in software engineering and forms the basis for several quality assessment models, including Hierarchical Software Quality Assurance (HSQA), implemented in tools like PIQUE~\cite{b15} and incorporated into frameworks such as the Consortium for IT Software Quality (CISQ) reliability measures~\cite{b14}. Static analysis plays a central role in these models by providing the inputs to HSQA models, such as PIQUE. By examining source code without executing it, static analysis tools can extract a wide range of software metrics, such as cyclomatic complexity, code duplication, coupling, and rule violations. Popular tools include commercial products like SonarQube, Coverity, and Fortify, as well as open-source tools such as PMD, Cppcheck and Bandit.

\looseness=-1
However, most static analysis tools rely on access to source code. In addition, many ISO/IEC 25010 characteristics and related metrics depend on constructs like class hierarchies, function definitions, or comments, which are unavailable in compiled binaries. Consequently, traditional quality models cannot be applied directly to malware in real-world scenarios, where source code is rarely accessible. This limitation highlights the need to reinterpret quality measurement for malicious software, focusing instead on features that can be extracted from binaries alone. Notably, ISO/IEC 25010 has also been applied in combination with complementary standards, demonstrating its flexibility as part of broader evaluation frameworks~\cite{b42}.

\section{Definitions of Malware Sophistication}
\label{sec:definitions}
Before presenting our quality assessment framework, it is important to establish definitions for terms such as ``sophistication", ``quality", ``complexity", and related concepts. These terms are often used inconsistently in the literature and are rarely defined in ways that facilitate their measurement. For this work, we adopt and adapt definitions that allow these concepts to be applied to malicious software, particularly when only binary artifacts are available. Table~\ref{tab:malware_definitions} summarizes the core terminology used throughout the paper. The definition of sophistication is central to this work: it emphasizes specialized knowledge and effort, distinguishing sophisticated malware from commodity variants.

We do not yet know how to directly measure the specialized knowledge and effort expended in malware development. However, we hypothesize that malware quality serves as an incomplete but partial measure of sophistication. The following section outlines how we propose to measure malware quality using inspiration from the ISO/IEC 25010 software quality standard.

\section{Application of ISO/IEC 25010 to Malware Quality Assessment}
\label{sec:iso}
\looseness=-1
Currently, no unified set of malware quality metrics exist that are equivalent to those used in traditional software engineering. To support consistent assessment of malware sophistication, it is first beneficial to establish a standardized framework of software quality applicable to malicious programs. To explore this goal, we examine the ISO/IEC 25010 standard, which defines a set of quality characteristics typically used to assess conventional software systems~\cite{b6}. Although it was originally designed for legitimate applications, these characteristics can provide a structured lens through which to evaluate malware, provided they are reinterpreted carefully. Table~\ref{tab:iso25010} summarizes each characteristic, with applicable characteristics listed first and excluded characteristics marked at the bottom. In the sections that follow, we first analyze the applicable characteristics in detail, then identify characteristics that are inapplicable or unmeasurable for malware.

\begin{table}[ht]
\centering
\caption{ISO/IEC 25010 Quality Characteristics}
\label{tab:iso25010}
\begin{tabular}{@{}p{1.15in}p{2.1in}@{}}
\toprule
\textbf{Characteristic} & \textbf{Definition} \\
\midrule
\multicolumn{2}{@{}l}{\textit{Applicable to Malware}} \\
\midrule
Reliability & Maintenance of performance under specified conditions \\[0.5ex]
Security & Protection of information and systems from unauthorized access \\[0.5ex]
Maintainability & Ease of modification for defects, improvements, or adaptation \\[0.5ex]
Flexibility & Adaptability to different requirements or environments \\[0.5ex]
\midrule
\multicolumn{2}{@{}l}{\textit{Excluded}} \\
\midrule
Functional Suitability & Degree to which functions meet specified requirements \\[0.5ex]
Performance Efficiency & Resource utilization relative to intended function \\[0.5ex]
Compatibility & Ability to operate within shared environments \\[0.5ex]
Interaction Capability & Effectiveness of user interaction with software \\[0.5ex]
Safety & Mitigation of risk to people, property, or environment \\
\bottomrule
\end{tabular}
\end{table}

\subsection{Reliability}
\looseness=-1
In malware, reliability translates to the ability of the sample to operate without faults, crashes, or unintended behavior during execution. Characteristics such as faultlessness and fault tolerance are particularly relevant when malware is designed to run persistently over extended periods.

\looseness=-1
From a binary analysis perspective, reliability may be inferred from structural indicators such as error-handling routines, redundant functionality, or fallback mechanisms. For example, multiple startup persistence mechanisms (e.g., registry keys, scheduled tasks, service installation) suggest design that tolerates removal attempts. The Nexe backdoor creates persistence via a scheduled task and checks mutex values to prevent redundant instances, illustrating fault tolerance~\cite{b80}.

\looseness=-1
Static analysis methods may increasingly support assessments of reliability. The Consortium for Information \& Software Quality (CISQ) Reliability Measures define 74 Common Weakness Enumerations (CWEs) intended to guide construction of reliable software in alignment with ISO/IEC 25010~\cite{b41}. Although this framework primarily targets source code, tools like CWE-Checker can detect a subset of reliability-related CWEs directly from compiled binaries.

\subsection{Security}
\looseness=-1
When applied to malware, security subcharacteristics are interpreted in reverse: sophisticated malware seeks to protect itself and its operators from discovery, analysis, and attribution. Confidentiality and integrity may be implemented through encryption, packing, or obfuscation to prevent reverse engineering~\cite{b16}. Non-repudiation and accountability are typically subverted through removal of logging or anonymized communication. Resistance, the most relevant subcharacteristic, refers to a program's ability to withstand external attacks, including antivirus evasion, anti-debugging, and sandbox detection.

\looseness=-1
Concrete examples illustrate these security traits. Nexe disables Windows security instrumentation by patching AMSI and ETW, encrypts network traffic, and obfuscates its payload~\cite{b80}. WikiLoader incorporates custom code and sandbox evasion to resist analysis~\cite{b82}. Sophisticated malware often incorporates C2 protocols that resist disruption and complicate attribution~\cite{b77}, using encryption or domain generation algorithms to evade detection.

\subsection{Maintainability}
\looseness=-1
Code reuse is a particularly telling signal. Shared codebases or modular architectures suggest efforts to develop and maintain malware at scale. Techniques like fuzzy hashing and function similarity analysis can identify commonalities across samples, revealing modular design or iterative versioning. Lehman's law observes that software tends to become more complex over time~\cite{b58}, and in malware, rising complexity may indicate ongoing development~\cite{b56,b57}.

\looseness=-1
Although maintainability is traditionally measured using source code metrics, several indicators can be inferred from binaries. Cyclomatic complexity can be approximated from disassembled code~\cite{b43}. However, some indicators may be distorted by obfuscation; file size may be artificially inflated through dead code insertion or padding~\cite{b60}. Additionally, maintainability features can conflict with an attacker's goal of avoiding attribution, as code reuse creates linkages that aid threat actor profiling~\cite{b78}.

\looseness=-1
Several campaigns reinforce these observations. The KANDYKORN malware exhibits a five-stage infection chain comprising multiple distinct components~\cite{b81}, while WikiLoader exhibits modularity and is believed to be sold to multiple actors~\cite{b82}.

\subsection{Flexibility}
\looseness=-1
In malware, flexibility can be understood as the ability to operate across varied system configurations, respond to changes in the target environment, or maintain functionality across a wide range of infected hosts.

\looseness=-1
Adaptability may be reflected in support for multiple operating systems or processor architectures, or in the ability to receive dynamic configuration updates from C2 servers. From a binary analysis perspective, flexibility indicators include conditional logic tied to OS versioning, abstraction layers for file paths or system APIs, and multiple payload stubs within a single executable. Malware may also alter behavior based on execution environment; for example, checks for sandboxing or virtualization may trigger modified execution paths~\cite{b44}.

\looseness=-1
A Lazarus-linked cross-platform infostealer illustrates this trait well. It deploys payloads compatible with Windows, Linux, and macOS, and includes functionality for file theft, keystroke logging, and crypto mining~\cite{b84}. Microsoft has similarly observed phishing campaigns that adapt delivery vectors and remote-access tools dynamically for each target~\cite{b85}.

\subsection{Excluded Characteristics}
\looseness=-1
Five ISO/IEC 25010 characteristics are either inapplicable to malware or not reliably measurable through static binary analysis.

\looseness=-1
\textbf{Functional suitability} requires determining whether software achieves its intended functionality correctly and completely. Without access to source code or specifications, this cannot be assessed for malware. Functional suitability also falls under ``Quality in Use," which requires user interaction~\cite{b36} that does not reflect real-world malware operation.

\looseness=-1
\textbf{Performance efficiency} can be partially inferred from static properties such as optimization level or avoidance of costly operations, but exact resource utilization requires dynamic analysis. While efficient resource usage may indicate mature development, this characteristic is most directly assessed at runtime.

\looseness=-1
\textbf{Compatibility} involves operating within shared environments without interference. However, malware strategy varies significantly: stealthy malware prioritizes interoperability~\cite{b38}, while destructive malware like HermeticWiper~\cite{b39} intentionally causes system crashes. This strategic variation complicates inclusion in a unified quality model.

\looseness=-1
\textbf{Interaction capability} is perspective-dependent, encompassing both attacker-side configuration and victim-side engagement. While some indicators (argument parsing, GUI components, localized strings) can be extracted from binaries, many malware samples lack interactive components entirely, making this characteristic difficult to generalize.

\looseness=-1
\textbf{Safety} attributes are largely inapplicable, as malware is inherently designed to operate without regard for user or system safety. Fail-safe behaviors in malware (such as terminating when a debugger is detected) serve stealth rather than safety and are more appropriately associated with reliability or security.

\section{Conclusion and Future Work}
\looseness=-1
This paper proposed a definition of sophistication as it pertains to malware. To address the lack of a formal malware quality framework, we adapted the ISO/IEC 25010 software quality standard by identifying characteristics applicable to malicious software. We focus on operationalizing Quality through reinterpretation of the ISO/IEC 25010 standard for static binary analysis.
\looseness=-1
To advance malware quality studies, further work is needed to explore how existing software quality frameworks can be adapted. The long-term goal is supporting implementation and validation of a system that quantifies malware sophistication. A practical challenge lies in aggregating heterogeneous tool outputs under a unified scoring hierarchy. It may be possible to assign learned weights to sub-characteristics using empirical data, but this requires a labeled corpus reflecting malware sophistication degrees. No such public dataset currently exists; future efforts must address this gap through expert annotation, consensus labeling, or semi-supervised learning.
\looseness=-1
A near-term priority is applying this framework to real malware samples to validate the proposed measurement process and assess whether the identified characteristics yield meaningful, consistent signals in practice. Once an operational model is established, statistical testing could confirm or challenge this assumption and further refine our understanding of adversarial software engineering.

\section*{Acknowledgements}
\looseness=-1
This work was supported by DHS S\&T at Sandia National Laboratories, a multi-mission laboratory managed and operated by National Technology \& Engineering Solutions of Sandia, LLC (NTESS), a wholly owned subsidiary of Honeywell International Inc., for the U.S. Department of Energy's National Nuclear Security Administration (DOE/NNSA) under contract DE-NA0003525. This written work is authored by an employee of NTESS. The employee, not NTESS, owns the right, title and interest in and to the written work and is responsible for its contents. Any subjective views or opinions that might be expressed in the written work do not necessarily represent the views of the U.S. Government. The publisher acknowledges that the U.S. Government retains a non-exclusive, paid-up, irrevocable, world-wide license to publish or reproduce the published form of this written work or allow others to do so, for U.S. Government purposes. The DOE will provide public access to results of federally sponsored research in accordance with the DOE Public Access Plan. 
The ChatGPT Large Language Model was used in this paper for spell-checking and minor grammatical enhancements. 

\balance

\vspace{12pt}

\end{document}